\documentstyle[prd,aps,epsfig,floats]{revtex}
\begin{document}
\draft
\wideabs{
\title{About the Functional Form of the Parisi Overlap Distribution 
for the Three-Dimensional Edwards-Anderson Ising Spin Glass}
\author{Bernd A. Berg$^1$, Alain Billoire$^2$ and Wolfhard Janke$^3$} 
\address{ (E-mails: berg@hep.fsu.edu, billoir@spht.saclay.cea.fr,
wolfhard.janke@itp.uni-leipzig.de)\\ 
$^1$Department of Physics, The Florida State University,
     Tallahassee, FL 32306, USA\\
$^2$CEA/Saclay, Service de Physique Th\'eorique, 
     91191 Gif-sur-Yvette, France\\
$^3$Institut f\"ur Theoretische Physik, Universit\"at Leipzig,
         04109 Leipzig, Germany}
\date{August 2, 2001}
\maketitle
\begin{abstract}
Recently, it has been conjectured that the statistics of extremes is
of relevance for a large class of correlated system. For certain
probability densities this predicts the 
characteristic large $x$ fall-off behavior $f(x)\sim\exp (-a\,e^x)$, 
$a>0$.  Using a multicanonical Monte Carlo technique, we have calculated 
the Parisi overlap distribution $P(q)$ for the three-dimensional 
Edward-Anderson Ising spin glass at and below the critical temperature, 
even where $P(q)$ is exponentially small. We find that a probability
distribution related to extreme order statistics gives an excellent 
description of $P(q)$ over about 80 orders of magnitude.
\end{abstract}
\pacs{PACS: 75.10Nr,75.40Mg,75.50Lk.}
}

\narrowtext

% \section{Introduction} \label{sec_intro}

The three-dimensional (3D) Edwards-Anderson\cite{EA75} Ising (EAI) 
spin-glass model is a prototype of a disordered system, for 
which conflicting constraints create a rough free energy landscape. 
Such systems are of importance for the understanding of a wide 
range of phenomena in physics, chemistry, biology and computer science.  
The overlap $q$ 
between two replicas of the EAI model serves as an order parameter. 
Its probability density $P(q)$ is, therefore, a quantity of central 
physical interest. More than twenty years ago, Parisi succeeded 
to calculate $P(q)$ in the mean-field approximation\cite{MPV87}. 
% which is elaborate due to replica symmetry breaking. 
However, for 3D 
physical systems the precise form of $P(q)$ in the spin-glass phase, 
and the very nature of this phase, have remained a subject of 
debate\cite{FiHu88,Yo97}. More recently connections between 
spin glasses and extreme order statistics\cite{Gu58,Ga87} were pointed 
out by Bouchaud and M\'ezard\cite{BoMe97}. Then, Bramwell et 
al.\cite{Br00} conjectured that the statistics of extremes, in 
particular Gumbel's first asymptote (introduced below), is of
relevance for a large class of correlated systems.  We have used a 
multicanonical Monte Carlo (MC) technique\cite{BeNe92,BeJa98} to calculate 
$P(q)$ numerically, even when it is exponentially small. At the critical
point a modification of Gumbel's first asymptote
gives a perfect description of the data over about 80 orders of 
magnitude and the agreement appears to continue below the critical
temperature. Although the detailed relationship between extreme order 
statistics and the EAI model remains to be understood, it is certainly 
quite rare that a physical formula has been tested over such a large
range.

The statistics of extremes was pioneered by Fr\'echet, Fisher and
Tippert, von Mises, and Gumbel. A standard result\cite{Gu58,Ga87},
due to Fisher and Tippert, Kawata, and Smirnow, is the universal 
distribution of the first, second, third,\dots smallest of a set of 
$N$ independent identically distributed random numbers. For an 
appropriate, exponential decay of the random number distribution 
their probability densities are given by
\begin{equation} \label{Gumbel}
f_a(x)\ =\ C_a\,\exp\left[\,a\,\left( x-e^x\,\right)\right]
\end{equation}
in the limit of large $N$. Here $x$ is a scaling variable, which 
shifts the maximum value of the probability density to zero, and 
$C_a=a^a/\Gamma(a)$ normalizes the integral over $f_a(x)$ to one. 
In Gumbel's book~\cite{Gu58} Eq.~(\ref{Gumbel}) is 
called the first asymptote, as it holds for the asymptotic extreme 
order statistics of the first of altogether three different 
universality classes of random number distributions. 
The exponent $a$ takes the
values $a=1,2,3,\dots $, corresponding, respectively, to the first, 
second, third,\dots smallest random number of the set. This holds 
independently of the details of the original random number distribution,
as long as one stays within the first universality class. 
In the last years a 
non-integer value of the exponent $a$ received also some attention. 
% Starting from an observation of universality between rare fluctuations 
% of turbulence in a coupled rotor model and critical phenomena in the 
% 2D XY model\cite{BHP98}, 
For the probability density of the magnetization of the 2D XY model
Bramwell et al.\cite{Br00,Br01} derived $a=\pi/2$ in the spin wave 
approximation and conjectured that this exponent describes, at least 
approximately, probability densities of a large class of correlated 
systems.
% , claimed to include (besides the mentioned 
% systems) the Ising model just below the critical temperature, 
% percolation models and some self-organized critical phenomena.

For disordered systems Bouchaud and M\'ezard\cite{BoMe97} noted that
a relationship to extreme order statistics is intuitively quite 
obvious. Namely, at low temperatures a disordered system will 
preferentially occupy its low energy states, which are random
variables due to the quenched exchange interactions of the system. 
Their investigation shows that Gumbel's first asymptote with 
$a=1$ corresponds to one step replica symmetry breaking, and
their conjecture of a relationship between 
extreme order statistics and disordered systems is certainly
far more general. This, and the possible description of a broad 
range of critical phenomena by the $a=\pi/2$ modification of Gumbel's 
first asymptote, has motivated us to analyze the overlap probability 
density of the EAI model at and below the critical point with respect 
to the large $x$ fall-off behavior of Eq.~(\ref{Gumbel}).

The energy function of the $J=\pm 1$ EAI 
spin-glass model is given by~\cite{EA75}
\begin{equation} \label{energy}
E = - \sum_{\langle ik \rangle} J_{ik}\, s_i s_k\ ,
\end{equation}
where the $s_i=\pm 1$ are the spins of the system and the sum is 
over the nearest-neighbor pairs of a cubic $L^3$ lattice with 
periodic boundary conditions. The coupling constants $J_{ik}$ are 
quenched random variables, which take on the values $\pm 1$, with 
equal probabilities. 
% {\it i.e.} the energy units are chosen such that the exchange 
% energies are $\pm 1$. 
A set of coupling constants defines a realization 
${\cal J}=\{ J_{ik}\}$ of the system. The two replica overlap 
(Parisi order parameter) is defined by 
\begin{equation} \label{q}
 q = {1\over L^3} \sum_{i=1}^{L^3} s_i^{(1)} s_i^{(2)}\ ,
\end{equation}
where the $s^{(1)}_i$ and $s^{(2)}_i$ are the spins of two copies 
(replica) of the realization ${\cal J}$ and the sum is over all sites. 
The overlap probability density is given by the average over the 
probability densities $P^{\cal J}(q)$ of all realizations
\begin{equation} \label{Pq}
P_L(q) = {1\over N_{\cal J}} \sum_{\cal J} P^{\cal J}(q)\ ,
\end{equation}
where $N_{\cal J}$ is the number of realizations used and $L$ is the 
lattice size. There is a long history of MC studies of this 
model\cite{Og85,BhYo88,KaYo96,In97,PaCa99,BBJ00,JBB01,Ba00},
which have led to a wealth of information. Here we introduce only 
two results: 1.~The model has a freezing transition at
a finite temperature, which according to the most recent 
estimates\cite{PaCa99,Ba00} is consistent with $T_c=1.14$,
where the temperature is given in dimensionless units such that the 
Boltzmann constant is one.
2.~References~\cite{KaYo96,In97,PaCa99,JBB01} reported results 
which were consistent with a Kosterlitz-Thouless~\cite{KT73} (KT) 
type line of critical points below $T_c$, quite similar to the 
2D XY model. In our context this is of interest in view of the 
description of this model by Eq.~(\ref{Gumbel}) with 
$a=\pi/2$~\cite{Br00}. Note, however, that one of the most recent EAI 
investigations\cite{Ba00} claims to rule out the KT scenario. 

\begin{figure}[-t] \begin{center}
\epsfig{figure=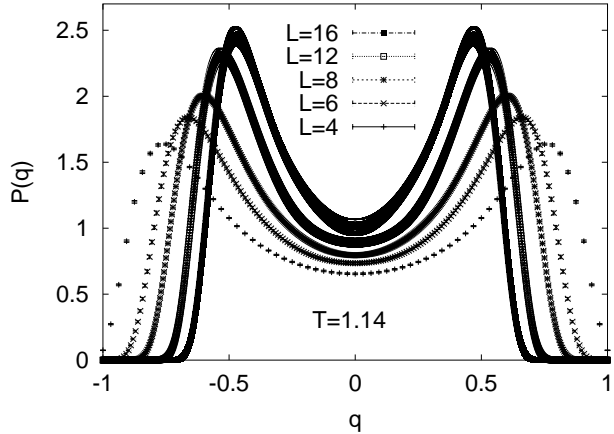,width=\columnwidth} \vspace*{-1mm}
\caption{Overlap probability densities $P_L(q)$ versus $q$ for the 
EAI model on $L^3$ lattices at the critical temperature. 
\label{fig_Pq3d} }
\end{center} \end{figure}

At $T=1.14$ we generated 8192 realizations for $L=4,\,6,\,8$, 512 
realizations for $L=12$ and 256 realization for $L=16$. 
Figure~\ref{fig_Pq3d} shows our normalized $P(q)$ probability densities.
% They rely on almost ??? thousand hours 375~MHz DEC Alpha processor 
% computer time.
Due to the MC method\cite{BeJa98} used, the error bars of neighboring 
entries are strongly correlated. This results in smooth curves of 
varying thickness, which represents the error. The pearl of these 
data are the tails of the distributions, which (for $L=16$) are 
accurate down to $10^{-160}$. This is achieved by simulating the 
system in a statistical ensemble for which the distribution of 
$q$-values is approximately flat, instead of using the Gibbs 
canonical ensemble.  After the simulation, results for the Gibbs 
ensemble are obtained through an exact re-weighting procedure. In 
this way computer simulations allow to probe much easier into the 
extremes of materials than real experiments.
Alongside with our data at the critical point, we analyze our
data from our simulations~\cite{BBJ00,JBB01} at $T=1$, below the
critical point. In that case we generated 8192 realizations for 
$L=4,\,6,\,8$ and 640 realizations for $L=12$. In the tails the 
data (for $L=12$) at $T=1$ are accurate down to $10^{-53}$.

\begin{figure}[-t] \begin{center}
\epsfig{figure=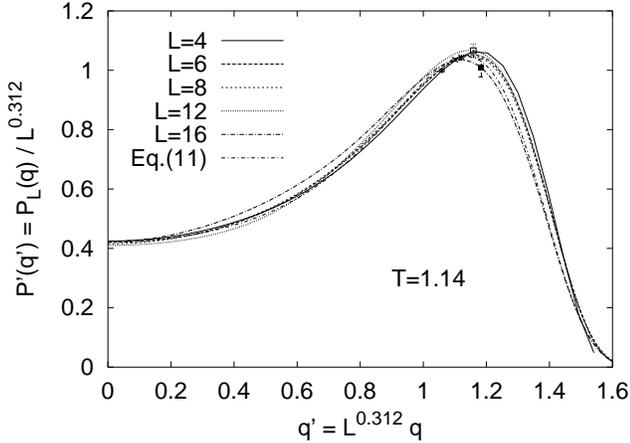,width=\columnwidth} \vspace*{-1mm}
\caption{Rescaled overlap probability densities for the EAI model 
on $L^3$ lattices at the critical temperature. \label{fig_Pp3d} }
\end{center} \end{figure}

We first ask the question whether, up to finite-size corrections, 
the probability densities depicted in Fig.~\ref{fig_Pq3d} scale.
A method to investigate this is to plot $\sigma_L\,P_L(q)$ versus 
$(q-\hat{q}_L)/\sigma_L$, where $\hat{q}_L$ is the mean value of 
$q$ with respect to the distribution $P_L(q)$ and $\sigma_L$ is 
its standard deviation (here $\hat{q}_L=0$ because the $P_L(q)$ 
are even functions). Visual inspection shows that the data
scale indeed and we proceed to fit the standard deviations to
the two parameter form $\sigma_L=c_1\,L^{-\beta/\nu}$ to obtain
\begin{eqnarray} \label{beta_nu1}
 {\beta\over\nu} &=& 0.312\,(4),\ Q=0.32\ {\rm for}\ T=1.14\,,
 ~~{\rm and} \\ \label{beta_nu2}
 {\beta\over\nu} &=& 0.230\,(4),\ Q=0.99\ {\rm for}\ T=1\,.
\end{eqnarray}
% Fits in STMC/ForProg/gfit_I3d on saclay wasa.
Here the numbers in parenthesis denote error bars with respect to 
the last digits and $Q$ is the goodness of fit. 
For $T=1.14$ we plot in Fig.~\ref{fig_Pp3d} 
$P'(q')=P_L(q)/L^{\beta/\nu}$ versus $q'=L^{\beta/\nu}\,q$ and see 
that the four probability densities collapse onto a single curve. 
To enlarge the scale, we restrict ourselves to the $q\ge 0$ range. 
The error bars of the lines in Fig.~\ref{fig_Pp3d} are up to 
multiplicative factors the error bars of Fig.~\ref{fig_Pq3d}. 
Not to obscure the agreement, we include only one representative 
error bar for each lattices size, $L=16,\,12,\,\dots$ from right 
to left in the Fig.~\ref{fig_Pp3d} (for $L\le 8$ they are barely
visible on the scale of the figure).
For our data at $T=1$ a similar analysis is already given 
in~\cite{JBB01}. The small discrepancy in the estimates of the 
critical exponent $\beta/\nu$ (0.255 in~\cite{JBB01} instead of 
0.230) is due to 
using  different methods of data analysis. Note that the error bars 
of the $\beta/\nu$ estimates (\ref{beta_nu1}) and (\ref{beta_nu2}) 
reflect only the fluctuations of our two parameter fit and additional 
(systematic) errors are expected from corrections to scaling.

Our aim is to relate the probability distribution of 
Fig.~\ref{fig_Pp3d} to the first asymptote of extreme order
statistics, Eq.~(\ref{Gumbel}). In the neighborhood of $x=0$
\begin{equation} \label{Taylor}
 x - e^x = - 1 - {1\over 2}\, x^2 + O\left( x^3 \right) 
\end{equation}
holds. To get the position of the maximum of $P'(q')$ right, we 
have to choose 
\begin{equation} \label{x}
 x = b\, (q' - q'_{\max})\ ,
\end{equation}
where $q'_{\max}$ is the $q'$ argument for which the probability 
density $P'(q')$ takes on its maximum value. We then have to find 
an exponent $a$ to reproduce the data for $x=b\,(q'-q'_{\max})>0$. 
For $x=b\,(q'-q'_{\max})<0$, however, the behavior (\ref{Gumbel}) 
cannot be quite correct. The reason is that the $x<0$ asymptotic 
behavior
\begin{equation} \label{negative_x}
\exp\left[\,a\,x\,\right]=\exp\left[\,a\,b\,(q'-q'_{\max})\,\right]
\end{equation}
predicts on a logarithmic scale a constant slope $a$ with decreasing $x$, 
while for the data of Fig.~\ref{fig_Pp3d} the slope levels off
and at $x=-b\,q'_{\max}$ we have 
\begin{equation} \label{derivative}
 \left. {d\, \ln [ P'(q')] \over d\ q'}\right|_{q'=0}\ =\ 0\ ,
\end{equation}
what is impossible with (\ref{negative_x}). A simple solution
is to replace the first $x$ on the right-hand
side of Eq.~(\ref{Gumbel}) by $c\,\tanh (x/c)$, where $c>0$ is
a constant. For small $x$ the Taylor expansion (\ref{Taylor}) still 
holds, while for large $|x|$ the hyperbolic tangent  function
$c\,\tanh (x/c)$ approaches 
quickly $\pm c$ (note that in the limit $c\to\infty$ the original 
form (\ref{Gumbel}) is recovered). For $x\to -\infty$ (practically
already at $q'=0$) the thus modified Gumbel distribution becomes 
constant. Therefore, the symmetric expression for $P'(q')$ is 
obtained by multiplying the above construction with its reflection 
about the $q'=0$ axis,
\begin{eqnarray} 
&P'(q')&\ =  C \times \label{mod_Gumbel} \\ \nonumber
&\exp &\left\{\,a\,\left[\,c\,\tanh \left(+{b\over c}\,(q'-q'_{\max})
\right) - e^{+b\,(q'-q'_{\max})}\right]\right\} \times \\ \nonumber
&\exp &\left\{\,a\,\left[\,c\,\tanh \left(-{b\over c}\,(q'+q'_{\max})
\right) - e^{-b\,(q'+q'_{\max})}\right]\right\}\ .
\end{eqnarray}
Of course, the important large $x$ behavior of Eq.~(\ref{Gumbel})
is not at all affected by our manipulations. 

The calculation of the parameters $a,\, b,\, c$ and $C$ is done by 
using the logarithm of Eq.~(\ref{mod_Gumbel}). The constants 
are essentially determined by the following parts of the 
distribution: $C$ by the height of the peak at $q'=q'_{\max}$
($q'_{\max}$ is now off the maximum location by a tiny shift which 
can be neglected; in the fits we used $q'_{\max}=1.135972$ for 
$T=1.14$ and $q'_{\max}=1.115056$ for $T=1$), $a$ and $b$ by the 
$q'>q'_{\max}$ tails of the distribution and $c$ by the value at 
$q'=0$. This allows to iterate from a reasonably close starting 
values to our final estimates
\begin{eqnarray} \label{a_par1}
a &=& 0.448\,(40)\ {\rm for}\ T=1.14\,, ~~{\rm and}
\\ \label{a_par2} a &=& 0.446\,(37)\ {\rm for}\ T=1\,.
\end{eqnarray}
%
% Obtained with rescale.f (options!), gfj_dat.f and gfj.f in
% berg/SPG/Pq_tail/gum3d and ../gum3dTc.
%
The non-universal coefficients are
$b=5.35\,(11)$, $c=3.37\,(41)$, $C=7.39\,(86)$ for $T=1.14$, and
$b=8.23\,(17)$, $c=4.48\,(43)$, $C=16.8\,(2.2)$ for $T=1$. The
error bars rely on a jackknife analysis.
% , to deal with the correlations in our $P(q)$ data.

\begin{figure}[-t] \begin{center}
\epsfig{figure=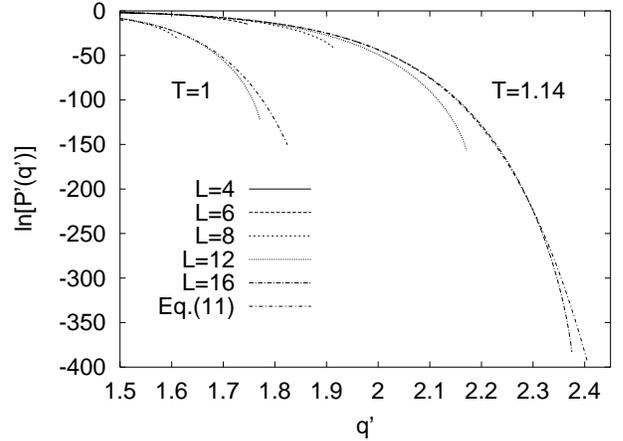,width=\columnwidth} \vspace*{-1mm}
\caption{Tails of the rescaled overlap probability densities of
Fig.~\ref{fig_Pp3d}: $\ln [P'(q')]$ versus $q'$. The error bars
are smaller or of the order of the thickness of the lines.
\label{fig_Ppln3d} }
\end{center} \end{figure}

For $T=1.14$ our best fit to Eq.~(\ref{mod_Gumbel}) is already
included in Fig.~\ref{fig_Pp3d}. In Fig.~\ref{fig_Ppln3d} we 
% continue where the resolution of Fig.~\ref{fig_Pp3d} ceases to
% exist and 
follow the tails of our distributions by plotting 
$\ln [P'(q')]$ versus $q'$ for $q'\ge 1.5$. Besides $T=1.14$, the
results for $T=1$ are also included in this figure. On the scale
of Fig.~\ref{fig_Ppln3d} the error bars are smaller than the
thickness of the lines, namely in the range 0.01 to 3.1 (exceeding
1.0 only for the largest $q'$ values of the $L=12$ and $16$ 
lattices). The reason is that our sampling is flat in $q$ and the
huge range of $\ln[P(q)]$ comes from subtracting logarithms of
(numerically exact) weight factors. 

Figure~\ref{fig_Ppln3d} exhibits the finite-size effect that for 
$q$ close to one the smaller lattices undershoot the larger ones. 
It is quite clear that something like this has to happen, because 
the data from each lattice size terminates at $q=1$, whereas 
Eq.~(\ref{mod_Gumbel}) has no corresponding singularity. 
When calculating our fit parameters, we take this into account by
restraining our use of data to 
$q'<2$, $\left(\ln[P(q')]>-43.4\right)$, for the $T=1.14$, $L=16$ 
lattice and to 
$q'<1.62$, $\left(\ln[P(q')]>-25.6\right)$, for the $T=1$, $L=12$ 
lattice. The agreement of our fits with those data stretches then 
over considerably larger ranges. Discrepancies of the $L=16$ data 
with the fit begin only around $\ln[P(q')]=-200$, and even this is 
only visible on a major enlargement of the scale of 
Fig.~\ref{fig_Ppln3d}. Discrepancies of the $T=1$, $L=12$ 
lattice with the fit are encountered around $\ln[P(q')]=-35$.
However, the $T=1.14$, $L=12$ data deviate already around 
$\ln[P(q')]=-10$ from the $L=16$ data. This appears possible,
because corrections to the $L^{0.312}$ scaling factor are not
traced by the accuracy of our data (in particular $L=16$ has
low statistics due to computer time limitations).  Therefore, 
it is not entirely clear whether the large range which we find 
for the agreement of the fit with our $L=16$ lattice is to some 
extent a statistical accident. Taking it at face value, we have the 
remarkable range of $200/\ln(10)\approx 87$ orders of magnitude. 
An actually smaller range would still be large enough to give us 
confidence that we are dealing with a true effect.

Our coefficient $a$ is far off from the 2D XY coefficient
of Bramwell et al.\cite{Br00}, $a=\pi/2$. 
This means that the EAI and the 2D XY models are certainly 
in quite different universality classes of extreme order statistics. 
However, the fact that both distributions can be described by it at 
all might help to explain the observed similarities. Our temperature 
$T=1$ is below the critical $T_c$, but with our lattice sizes it 
appears impossible to resolve the question, whether the here 
reported behavior reflects the existence of a critical line below 
$T_c$ or just the closeness of $T=1$ to $T_c$.
Before comparing to extreme order statistics, we~\cite{JBB01}
tried to fit the $q>q_{\max}$ tails of our distributions to the
theoretical predictions which have been made by Parisi and 
collaborators~\cite{Pa92,Pa93a,Pa93b} based on the replica mean-field 
approach. None of these fits was particularly good and even when 
pushing the adjustment of free parameters to their limits only small
parts of the tails of our distributions could be covered.

In summary, we have presented strong numerical evidence that the
Parisi overlap distribution of the EAI model can be described by
Eq.~(\ref{mod_Gumbel}). The detailed 
relationship between the EAI model and extreme order statistics
remains to be investigated and it is certainly a challenge to
extend the work of Bouchaud and M\'ezard\cite{BoMe97} to the more
involved scenarios of the replica theory. On the other hand, it
could be that replica symmetry breaking is not the driving mechanism 
of the EAI model phase transition and that our observations are rooted 
in general relations\cite{Br00} between correlated systems and extreme 
order statistics. 

\acknowledgments
We would like to thank Francois David and
Andr\'e Morel for discussions. This research 
was in part supported by the U.S. Department of Energy under contract
DE-FG-97ER41022. Most numerical simulations were performed on the T3E 
computers of the CEA in  Grenoble 
% under grant p526 
and of the NIC in J\"ulich.
% under grant hmz091. 
% Additional calculations were done on the T3E computer of ZIB in 
% Berlin and on Alpha workstations at the FSU. 
% We thank all institutions for their support.

\end{document}